\documentclass[twocolumn,pre,showpacs,superscriptaddress]{revtex4}
\pdfoutput=1 
\usepackage{graphicx}%
\usepackage{color} 
\usepackage{transparent}
\usepackage{psfrag}
\usepackage{dcolumn}
\usepackage{amsmath}
\usepackage{amssymb}
\usepackage{latexsym}
\usepackage{hyperref} 
\usepackage{booktabs}
\usepackage{latexsym}
\begin{document}

\def\bea{\begin{eqnarray}}
\def\eea{\end{eqnarray}}
\def\beq{\begin{equation}}
\def\eeq{\end{equation}}
\def\f{\frac}
\def\k{\kappa}
\def\e{\epsilon}
\def\ve{\varepsilon}
\def\be{\beta}
\def\D{\Delta}
\def\th{\theta}
\def\t{\tau}
\def\a{\alpha}
\def\J{{\cal J}}

\def\cDa{{\cal D}[X]}
\def\cDd{{\cal D}[X^\dagger]}
\def\cL{{\cal L}}
\def\cLo{{\cal L}_0}
\def\cLa{{\cal L}_1}

\def\Re{{\rm Re}}
\def\sj{\sum_{j=1}^2}
\def\rk{\rho^{ (k) }}
\def\rek{\rho^{ (1) }}
\def\cek{C^{ (1) }}
\def\rz{\rho^{ (0) }}
\def\rt{\rho^{ (2) }}
\def\rtb{\bar \rho^{ (2) }}
\def\trk{\tilde\rho^{ (k) }}
\def\trek{\tilde\rho^{ (1) }}
\def\trz{\tilde\rho^{ (0) }}
\def\trt{\tilde\rho^{ (2) }}
\def\r{\rho}
\def\tD{\tilde {D}}

\def\s{\sigma}
\def\kb{k_B}
\def\F{{\cal F}}
\def\la{\langle}
\def\ra{\rangle}
\def\nn{\nonumber}
\def\up{\uparrow}
\def\dn{\downarrow}
\def\S{\Sigma}
\def\dg{\dagger}
\def\d{\delta}
\def\p{\partial}
\def\l{\lambda}
\def\le{\left}
\def\ri{\right}
\def\L{\Lambda}
\def\G{\Gamma}
\def\o{\Omega}
\def\w{\omega}
\def\g{\gamma}

\def\jv{ {\bf j}}
\def\jr{ {\bf j}_r}
\def\jd{ {\bf j}_d}
\def\noi{\noindent}
\def\a{\alpha}
\def\d{\delta}
\def\p{\partial} 

\def\H{ {\bf H}}
\def\He{{\bf H_e}}
\def\h{{\bf h}}
\def\m{{\bf m}}
\def\hth{h_{\theta}}

\def\la{\langle}
\def\ra{\rangle}
\def\e{\epsilon}
\def\n{\eta}
\def\g{\gamma}
\def\break#1{\pagebreak \vspace*{#1}}
\def\hf{\frac{1}{2}}

\title{Macrospin in external magnetic field: Entropy production and fluctuation theorems} 
\author{Swarnali Bandopadhyay}
\email{swarnalib@tifrh.res.in}
\affiliation{TIFR Centre for Interdisciplinary Sciences, 
21 Brundavan Colony, Narsingi, Hyderabad 500075, Telengana, India
}
\author{Debasish Chaudhuri}
\email{debc@iith.ac.in}
\affiliation{Indian Institute of Technology Hyderabad,
Yeddumailaram 502205, Telengana, India
}
\author{A. M. Jayannavar}
\email{jayan@iopb.res.in}
\affiliation{Institute of Physics, Sachivalaya Marg, Bhubaneswar 751005, India
}
\date{\today}

\begin{abstract}
We consider stochastic rotational dynamics of a macrospin at a constant temperature, in presence of an external magnetic field.  Starting from the appropriate Langevin equation which contains multiplicative noise, we calculate entropy production (EP) along stochastic trajectories, and obtain fluctuation theorems.  The system remains inherently out of equilibrium due to a spin torque supporting azimuthal current, leading to an excess EP apart from the EP due to heat dissipation.  The anomaly may be removed using a redefinition of dissipated heat and stochastic work done.  Using numerical simulations, we obtain distribution functions for entropy production along stochastic trajectories to find good agreement with the detailed fluctuation theorem.   
\end{abstract}


\maketitle
\section{Introduction}
With miniaturization of memory devices like the magnetic read head and random access memory, thermal fluctuations start to play non-trivial role in their performance, e.g., by activating magnetization reversal of ferromagnetic clusters~\cite{Parkin2000}. 
The impact of thermal noise is stronger in smaller devices~\cite{Blanter2000,Jr1963}, with the relative intensity being inversely proportional to system size. 
The thermally induced magnetization fluctuations will act as a fundamental limit to the performance of submicron magnetoresistive devices. 
Thus, even from the application perspective, it becomes crucial to understand the impact of  thermal fluctuations, in order to reliably use small magnetic devices~\cite{Tserkovnyak2001,Foros2005,Foros2007,Bandopadhyay2011,Covington}. The simplest component of such magnetic devices is  a single magnetic domain, a macrospin.  

Thus it is interesting to understand {\em stochastic} thermodynamic properties of a macrospin under strong thermal fluctuations~\cite{Utsumi2015}. 
Stochastic counterparts of thermodynamic observables, e.g., energy, work, heat, and entropy characterizes stochastic trajectories in phase space. Statistical averages of these quantities lead to the corresponding macroscopic thermodynamic observables~\cite{Sekimoto1998,Jarzynski2011,Seifert2012}. 
Several equalities involving the stochastic observables
have been derived in the last two decades~\cite{Jarzynski1997,Baiesi2009,Baiesi2010,Hummer2010,Kurchan2007,Narayan2004,Jayannavar2007,Lahiri2014a}. 
It is shown that negative entropy producing trajectories do occur, but probability of them remains exponentially suppressed with respect to the positive entropy producing trajectories. The corresponding equality is known as the detailed fluctuation theorem (DFT)~\cite{Evans1993,Gallavotti1995,Lebowitz1999,Crooks1999,Seifert2005,Saha2009,Lahiri2009,Sahoo2011}.  
A related integral fluctuation theorem (IFT), and the Jarzynski equality that expresses equilibrium free energy difference in terms of non-equilibrium work done were also derived~\cite{Jarzynski1997,Crooks1999}. 
Many of these theorems have been verified against experiments on colloids and granular 
matter~\cite{Wang2002,Blickle2006,Speck2007,Joubaud2012}, and successfully used to obtain free energy landscape of bio-polymers like RNA~\cite{Liphardt2002,Collin2005}. 

Recently, stochastic thermodynamics has been extended to describe active Brownian particles that derives their motion using internal energy source or ambient 
fuel~\cite{Hayashi2010,Seifert2011,Ganguly2013,Chaudhuri2014}.  Several active particle dynamics are describable in terms of non-linear velocity dependent forces having odd parity under time-reversal. 
For these systems, it was shown that the entropy production (EP) in the environment has excess contributions, apart from the contribution from dissipated heat~\cite{Ganguly2013,Chaudhuri2014}. 
This excess EP is shown to be related to the time reversal symmetry breaking due to the presence of non-linear velocity dependent force, and potential energy of interaction 
or trapping, together~\cite{Chaudhuri2014}.  

Using a single Ising spin undergoing Glauber dynamics, work distribution functions have been studied earlier, 
under various protocols of time dependent variation of magnetic field~\cite{Marathe2005}. 
If the external field is time-independent,  such a system reaches an equilibrium Boltzmann distribution when coupled to a heat bath. 
Thus to extract non-equilibrium entropy production, it is imperative to impose a time-dependent field on such a system.
We instead focus on the three dimensional stochastic motion of magnetization 
of a sufficiently {\em small} single magnetic domain, a macrospin, coupled to a Langevin heat bath, under external magnetic field~\cite{Jayannavar1991,Seshadri1982}. 
One fundamental difference in dynamics of this system with respect to the Ising spin is, the presence of a directed precessional  motion around the external
field, even if the field is time independent. Thus the system in the presence of a magnetic field is intrinsically out of equilibrium. 
Due to the small system size the dynamics of such a {macrospin} is strongly influenced by thermal noise. 

We use the Langevin and corresponding Fokker-Planck equations describing stochastic motion of a macrospin under external magnetic field. The appropriate Langevin equation contains multiplicative noise, and the external magnetic field gives rise to a spin torque. The Fokker-Planck equation is used to calculate rate of entropy change in the system, which has two terms, one is the EP due to non-equilibrium processes in the system, and another term gives entropy flux to the reservoir. The derivation clearly shows that the total average EP in the system and reservoir is non-negative, as required by the second law of thermodynamics. 
Using probabilities of stochastic trajectories we derive fluctuation theorems for stochastic EP. The corresponding expression of EP in the reservoir, $\D s_r$, depends on  the choice of time-reversed conjugate trajectories, only one of which is consistent with the Fokker-Planck equation. A direct derivation
of the stochastic energy balance shows that the EP in the environment has an excess contribution, apart from the stochastic version of Clausius entropy associated with heat dissipation.  However, this anomaly may be resolved by redefining both the dissipated heat $-\D Q \equiv T\D s_r$ with $T$ denoting the temperature of the reservoir, and the stochastic work done on the system, keeping the expression of change in internal energy unchanged.  Within the redefined form, these two terms contain contribution from rotational work done due to torque.  We propose experiments to separately measure heat dissipation and reservoir EP  in the presence of  spin torque, to test the relation between the two. Such measurements will help to understand stochastic EP better.

\section{Model}
The deterministic dynamics of a macrospin having magnetization $\m$ under time-dependent magnetic field $\H(t)$ is described by $\dot \m = \g \m \times \H(t)$, where $\dot\m = d\m/dt$, and $\g$ denotes the gyromagnetic ratio. However, the macrospin is not isolated, and its dynamics gets affected by the surrounding medium. In the simplest idealization, the effect of medium could be incorporated in terms of a   stochastic force conjugate to magnetization, a magnetic field $\h(t)$ varying randomly with time~\cite{Kubo1962}. 
Thus the effective dynamics becomes  $\dot \m = \g \m \times (\H(t) + \h(t))$. If $\h(t)$ is modeled as
a Gaussian noise, one obtains the well-known Bloch-equation~\cite{Kubo1962,Seshadri1982}. However, within such a description, the expectation value of magnetization relaxes to zero even in presence of a constant external magnetic field. This situation corresponds to the case of infinite temperature. In a finite temperature Langevin dynamics, the stochastic force is necessarily coupled to a frictional dissipation, obeying fluctuation-dissipation theorem.  
The corresponding Langevin dynamics has the Landau-Lifshitz-Gilbert (LLG) form~\cite{Jr1963, Kubo1970} 
\begin{equation}
\dot \m=\gamma \, \m \,\times\, \left[\H+\h(t)-\eta\dot \m \right].
\label{LLG1}
\end{equation}
 where $\eta$ is the the Gilbert damping coefficient~\cite{Gilbert1955}. The stochastic magnetic field obeys Gaussian statistics with 
 \bea
 \la \h (t) \ra  &=& 0, \nn\\ 
 \la \h (t) \otimes \h(t')\ra &=& 2  D_0 {\bf 1} \d(t-t')
 \label{noise}
 \eea
 where ${\bf 1}$ denotes the identity matrix, $D_0=\eta \kb T/V$ with $T$ denoting the temperature, $\kb$ the Boltzmann constant, and $V$ the volume of the magnetic particle. To get an assessment of the strength of the stochastic field $\h (t)$, let us evaluate it for a magnetic particle of volume $V= (10 \,{\rm nm})^3$ at room temperature, over a timespan  of $1$\,ps. The corresponding strength is $\sim 10$\,mT. Compare it with the field strength near a magnetic tape $\sim 1\,\mu$T, or, earth's magnetic field, which varies between $25$ to $65\,\mu$T. However, the very stochastic nature of $\h (t)$ ensures that its integrated impact over longer time-spans reduces in magnitude, leaving fluctuations in the orientations of $\m$. 
Note that the LLG equation conserves the amplitude of  magnetization $m=|\m|$, $d (m^2)/dt = 0$. This is a valid approximation for macrospins made of Ferromagnetic material like Fe or Co with Curie temperature $T_c \sim 10^3$K, three to four times higher than 
the room temperature, leading to negligible fluctuations in $m$~\cite{Blundell2001}. 

The above mentioned phenomenology can be derived from microscopic equations of motion (classical or quantum), by using the Zwanzig formalism of coupling the system dynamics with that of a heat bath composed of infinitely large degrees of freedom,  and then integrating out the heat bath degrees of freedom from the system equation of motion, and finally using a Markovian  approximation~\cite{Zwanzig2001}. This was achieved in two earlier studies using two different kinds of heat baths. In Ref.~\cite{Seshadri1982}, the surrounding environment of the magnetization was assumed to be composed of spins. A bilinear coupling of the system magnetization with environment spins led to the appropriate Langevin dynamics, which is equivalent to Eq.(\ref{LLG1}). In Ref.~\cite{Jayannavar1991}, dynamics of a single magnetic particle was coupled to a Harmonic oscillator heat bath, using a bilinear coupling between the magnetization and displacements of the oscillators. After integrating out the oscillator degrees of freedom one obtains the corresponding Langevin dynamics. This reduces to the above mentioned LLG form of Eq.(\ref{LLG1}) within the Markovian approximation. This approximation makes the memory kernel a constant $\eta$, the Gilbert damping coefficient, and gives rise to a delta-function correlated stochastic magnetic field $\h (t)$, having a mean value $\la \h (t) \ra =0$, and obeying the fluctuation-dissipation relation  $\la \h (t) \otimes \h(t')\ra = 2  D_0 {\bf 1} \d(t-t')$ with $D_0=\eta \kb T/V$ as expressed  in Eq.(\ref{noise})~\cite{Jayannavar1991}.

Thus,  to describe the stochastic dynamics of a macrospin having magnetization $\m$ under an external field $\H(t)$, we use the LLG equation given in Eq.(\ref{LLG1}).
The strength of stochastic noise in this equation depends on the magnetization $\m (t)$, i.e., the noise is multiplicative.
In the above equation, $\H$ denotes the conservative field $\H=-\p G/\p \m$, where $G = -\m.\H$ is the Gibb's free energy per unit volume.
The macrospin undergoes a relaxation dynamics in the Langevin heat bath, settling into an average unidirectional precession around the field $\H$, conserving the amplitude $m$. 

\begin{figure}[t]
\begin{center}
\includegraphics[width=8cm]{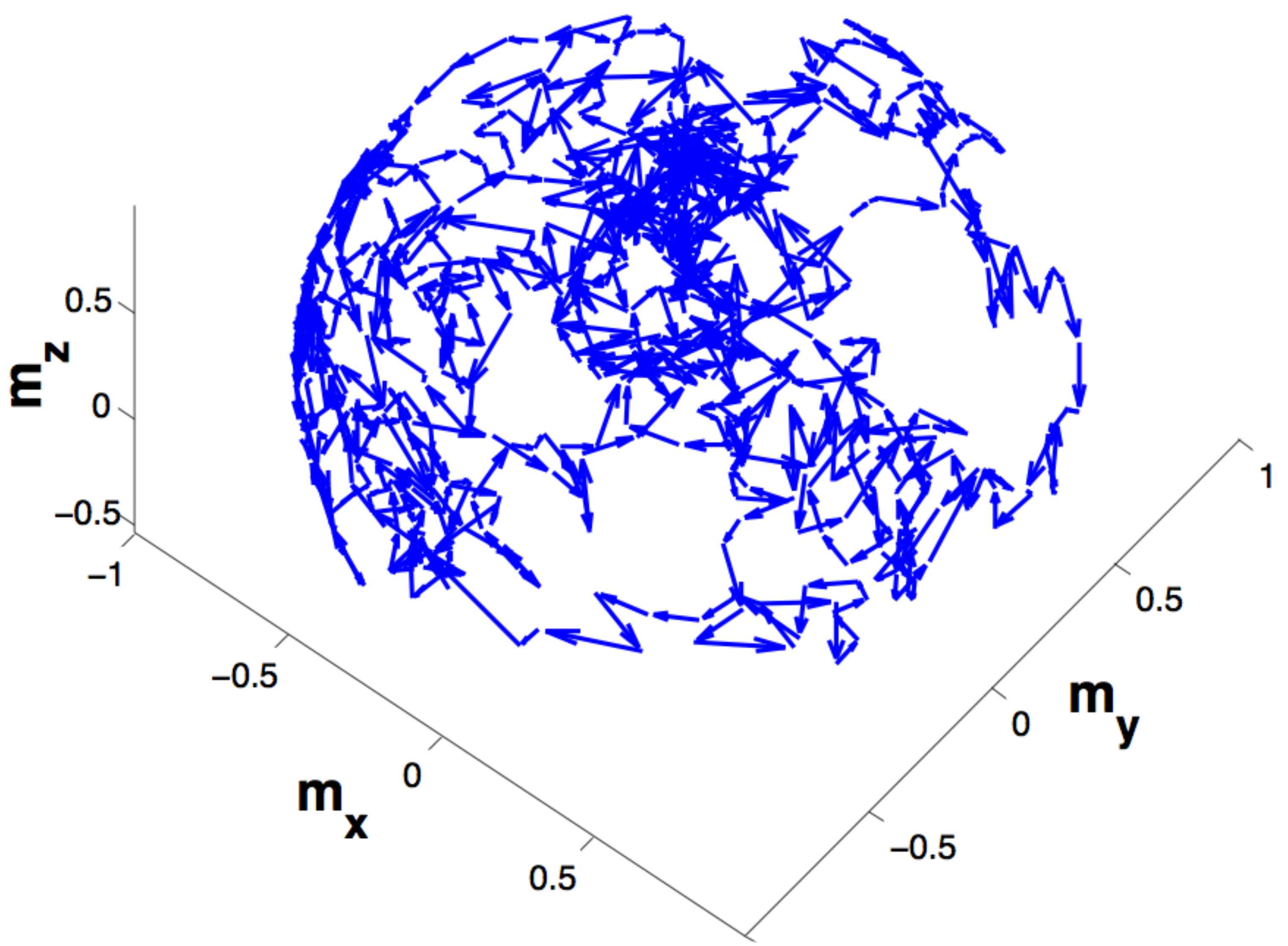}
\caption{(Color online) A typical  trajectory of magnetization $\m$ in presence of a magnetic field $\H = H \hat z$ with $H=1$ and a Langevin heat bath at temperature $\kb T=1$, obtained from numerical simulations. The arrow heads denote the direction of motion.}
\label{conf}
\end{center}
\end{figure}

This angular dynamics of the instantaneous orientation of magnetization $\m$ may be represented in terms of polar and azimuthal angles $(\theta (t),\phi (t))$ on the surface of a sphere of a fixed radius 
$m$. Using this description, the LLG equation can be expressed as
\bea
\dot\th &=& h'\, m( H_\th+ h_\th) - g' m (\sin\th)^{-1} (H'_\phi +  h_\phi) \nn\\
\sin\th \, \dot \phi &=& g'  m(H_\th +   h_\th) + h' m (\sin\th )^{-1}  (H'_\phi +  h_\phi),\nn\\
\label{llg2}
\eea
where
\bea 
 g' = \f{1/\g m}{ (1/\g^2)+\eta^2 m^2 },\, h'=\f{\eta}{(1/\g^2)+\eta^2 m^2}, \nn
\eea
with $\dot\th=\p_t \th$, $\dot\phi=\p_t\phi$, and
$H_\th = -({1}/{m})\p_\th G$, $H'_\phi \equiv H_\phi \sin\th = -({1}/{m})\p_\phi G$ so that one can express $\H = \hat \th H_\th + \hat \phi H_\phi$ in spherical polar coordinates.
The components of stochastic field are given by
$h_\th = h_x \cos\th \cos\phi+ h_y \cos\th \sin\phi - h_z\sin\th$, 
$h_\phi = -h_x\sin\th \sin\phi + h_y \sin\th \cos\phi$. 
Note that the radial component 
$h_r = h_x \sin\th \cos\phi + h_y \sin\th \sin\phi + h_z \cos\th$
does not appear in the angular motion of magnetization.
A typical steady state trajectory obtained from numerical simulations is shown in Fig.\ref{conf}. The details of the simulations will be discussed in Sec.~\ref{simu}.

{ In a recent study~\cite{Aron2014}, it has been shown that the form of Fokker-Planck equation derived from the LLG equation is independent of the choice of stochastic calculus --
Ito, Stratonovich or a post-point discretization scheme~\cite{Lau2007a, vanKampen1992}.
The Fokker-Planck equation was originally derived in Ref.~\cite{Jr1963} using the Stratonovich convention, which
we use  throughout this paper.}

\section{Fokker-Planck equation and entropy production}
A statistical ensemble of magnetization orientations  can be described by the surface probability density $P\le(\theta,\phi,t \ri)$. 
The corresponding Fokker-Planck equation is expressed as 
\bea
\p_t P = -{\bf \nabla}_{\o}. {\bf J}_\o,~~  {\bf J}_\o =\hat\th J_\th + \hat \phi J_\phi  
\label{fpeq}
\eea
where the two-dimensional divergence on the surface of the unit  sphere 
${\bf \nabla}_{\o} . {\bf J}_\o = \f{1}{\sin \th} \p_\th (\sin \th J_\th ) + \f{1}{\sin \th} \p_\phi J_\phi$, $\o$ denotes a 
solid angle. 
The two components of dissipative current are given by~\cite{Jr1963}
\bea
J_\th &=&  m [h' H_\th - g' H_\phi] P - k' \p_\th P \nn\\    
 J_\phi &=&   m [ g' H_\th + h' H_\phi] P - k' (\sin\th)^{-1} \p_\phi P. 
 \label{current}
\eea
In the above relations, $h'$ and $g'$ play the role of mobility, and $k'$ plays the role of diffusivity for angular dynamics.  
These mobility and diffusivity coefficients obey Einstein like relation 
$k'=D_0 m^2 (h'^2 + g'^2) = \kb T \, [\eta \g^2/V(1+\g^2 \eta^2 m^2)]$. 
 
The  non-equilibrium Gibbs entropy is given by~\cite{Crooks1999, Seifert2005} 
$$S = -\kb \int d\o\, P(\th,\phi,t) \ln P(\th,\phi,t) = \la -\kb \ln P \ra,$$
where $\int d \o =\int \sin\th\, d\th\, d\phi\,$ denotes integration over the phase space, which in this case is all possible solid angles, and
$\la \dots \ra$ denotes statistical average. 
Note that this definition of $S$ is equivalent to the Shanon information entropy associated with any probability distribution~\cite{Shanon1948, Kardar2007}.
 The generic paradigm of  Maxwell's daemon paradox~\cite{Szilard1929} helped building the connection between 
Shanon's information entropy and thermodynamic entropy~\cite{Maruyama2009,Mandal2012,Leff2014}. Recent experiments verified
that it is possible to convert information to free energy~\cite{Toyabe2010}. 
The direct relation between information entropy and physical entropy was exemplified by 
Landauer's principle linking the minimum heat dissipation associated with erasure of one bit of information as $\kb T \ln 2$~\cite{Landauer1961}.
This has been recently verified experimentally~\cite{Berut2012}. 
Thus the above definition of  entropy $S$ has a much wider scope going beyond equilibrium physics. This includes non-equilibrium systems as well.
The stochastic entropy of a microscopic state of the system is $s(\th,\phi,t) = -\kb \ln P(\th,\phi,t)$,
with the entropy of the ensemble  $S = \la s \ra$.
One can express the rate of change in stochastic entropy as 
\bea
\f{\dot s}{\kb} &=& -\f{\p_t P}{P} - \f{\p_\th P}{P} \dot \th - \f{\p_\phi P}{P} \dot \phi \nn\\
&=& -\f{\p_t P}{P} + \f{J_\th \dot \th + J_\phi \sin\th\, \dot\phi}{k' P} - \f{\dot s_r}{\kb}, 
\label{sdot}
\eea
where, 
\begin{align}
&\f{\dot s_r}{\kb} =
\f{m}{k'} \le[ h' (H_\th \dot\th + H_\phi \sin\th \dot\phi) + g' (H_\th \sin \th \dot \phi - H_\phi \dot \th)\ri]\nn\\
&= V \f{m}{\kb T} \le[ (H_\th \dot\th + H_\phi \sin\th \dot\phi) + \f{1}{m\eta\g} (H_\th \sin \th \dot \phi - H_\phi \dot \th)\ri].
\label{srdot}
\end{align}
The first step in Eq.(\ref{sdot}) identifies the explicit and implicit time dependences. The second step is obtained by using Eq.(\ref{current}) to replace $\p_\th P$ and
$\p_\phi P$. 
In obtaining the second step in Eq.(\ref{srdot}) we used  $k'=D_0 m^2 (h'^2 + g'^2)$, and the expressions for $h'$ and $g'$, along with the identity $D_0 =\eta \kb T/V$. 

At this point, we focus on the interpretation of the second term on right hand side of Eq.(\ref{sdot}), 
${\cal T} = [J_\th \dot \th + J_\phi \sin\th\, \dot\phi]/k' P$. Let us perform a two step averaging on Eq.(\ref{sdot}): (i)~over trajectories and (ii)~over 
the ensemble of all possible solid angles $\o$ with probability $P(\o,t)$. The trajectory average of the components of angular velocity leads to
$\la \dot \th |\, \th, \phi, t\ra = J_\th/P$ and   $\la \sin\th\, \dot \phi |\, \th, \phi, t\ra = J_\phi/P$~\cite{Seifert2005}. Thus after averaging over trajectories, ${\cal T}$ can be replaced by the expression
$\bar {\cal T} = (J_\th^2 +J_\phi^2)/k'P^2$. 
Now, to perform averaging over the phase space probability $P(\o,t)$, we multiply Eq.(\ref{sdot}) throughout by $P(\o,t)$ and integrate over $\o$.
The conservation of probability $\int d\o P(\o,t)=1$ leads to $\int d\o \p_t P(\o,t) = 0$. Thus one obtains the average EP in the system
\bea
\dot S \equiv \la \dot s\ra = {\kb}\int d\o  \f{ J_\th^2 + J_\phi^2}{k' P}  - \la \dot s_r \ra \equiv  \Pi - \la \dot s_r \ra,
\eea
where, 
$$\Pi \equiv \kb \la \bar {\cal T} \ra = \kb \int d\o  \f{ J_\th^2 + J_\phi^2}{k'P} $$
is the EP due to irreversible non-equilibrium processes occurring in the system quantified by $J_\th$ and $J_\phi$, and $\la \dot s_r \ra$ is the entropy flux from the system to the surrounding environment.
This second term quantifies the EP in the environment.
Thus the total EP in the combined system and environment,
$$\dot S_t = \dot S +  \la \dot s_r \ra = \Pi \geq 0,$$ 
 obeys the second law of thermodynamics.
At non-equilibrium steady states, $\Pi = \la \dot s_r \ra$, i.e., whatever entropy is produced in the system flows out to the environment. 
The preceding analysis shows that $\dot s_r$ is the stochastic EP in the environment.    
Note that the expression of $\dot s_r$, as given by Eq.(\ref{srdot}),  consists of terms having dimensions of torque times angular velocity, similar to dissipated work that one obtains from usual Langevin dynamics of particles  moving in a medium.

As we show in the following section, the average change in the total stochastic entropy $s_t = s + s_r$, calculated over a finite time interval $\t_0$,  can be interpreted as
the Kullback-Leibler divergence between the  time forward and time-reversed distribution of trajectories. Thus,  $\D s_t = \int_0^{\t_0} dt\, \dot s_t$ quantifies the
breakdown of time-reversal symmetry.

\section{Fluctuation theorems}
Now we proceed to derive EP along stochastic trajectories. 
Physically, EP characterizes the irreversibility of a trajectory. 
Consider the time evolution of a macrospin from $t=0$ to $\t_0$ through a path $X = [\th(t), \phi(t), \H(t)]$, assuming for the moment a time-dependent protocol of controlling $\H(t)$. 
Let us divide the path into $i=1,2,\dots,N$ segments of time-interval $\d t$ with $N \d t = \t_0$.
The transition probability $p_i^+ (\th', \phi', t+\d t | \th, \phi,t)$ on $i$-th infinitesimal segment is governed by the Gaussian random noise 
$P({\bf h}_i) = (\d t/4\pi D_0)^{1/2} \exp(-\d t\, {\bf h}_i^2/4 D_0)$ where ${\bf h}_i^2$ 
is calculated at $i$-th instant. 
Denoting Eq.(\ref{llg2}) as  
$\dot \th = \Theta(\th, \phi, \H)$ and $\dot \phi = \Phi(\th,\phi, \H)$, %
the transition probability on $i$-th segment
$p_i^+ = \J_i \la \d(\dot \th_i -\Theta_i) \d(\dot \phi_i - \Phi_i) \ra =  \J_i  \int d {\bf h}_i P({\bf h}_i) \d(\dot \th_i -\Theta_i) \d(\dot \phi_i - \Phi_i)$. 
The  Jacobian of transformation $\J_i={\rm det}[\p(h_{x_i}, h_{y_i}, h_{z_i})/\p(m_i, \th_i, \phi_i) ]_{m_i = {\rm constant}}$.
The probability of a full trajectory is ${\cal P}_+ = \prod_{i=1}^N p_i^+$.

It is possible to choose conjugate dynamics and trajectories in several ways, each of which will give rise to a new { entropy-like} quantity obeying detailed and integral 
fluctuation theorems, as shown in Appendix-\ref{Ap-A}. 
This fact has been discussed in literature and questions have been posed as to which choice would be physically meaningful~\cite{Seifert2012,Turitsyn2007,Speck2008}. 
For example, consider entropy production in soft matter system in external shear flow~\cite{Turitsyn2007,Speck2008}. While considering conjugate trajectories, one possibility
is to assume that the external flow does not change direction with respect to the time-forward trajectories, treating the flow as a quantity similar to external force~\cite{Turitsyn2007}. 
This gives rise to an expression of { entropy}, which obeys fluctuation theorems~\cite{Turitsyn2007}. On the other hand, one can also extend the operation of time-reversal to the 
particles of fluid. Then for conjugate trajectories fluid flow changes sign. This leads to a different expression of entropy, again obeying fluctuation theorems. It was argued in Ref.~\cite{Speck2008} that this second choice is physically more appealing. It is clear that the conjugate dynamics has to be chosen carefully to produce physically meaningful expression of entropy~\cite{Spinney2012}.

For the present problem, the probability distribution of micro-states $P(\th,\phi,t)$ evolve via the Fokker-Planck equation, from which we already obtained EP.  Calculation of EP using probabilities of time-forward and conjugate trajectories should agree with the outcome of the Fokker-Planck equation. Thus care has to be taken so that the choice of conjugate trajectories is consistent with the symmetries of the Fokker-Planck equation. 

The time-forward trajectory $X$ considered above has $\H (t)$ as the control parameter. The above guiding principle gives us a unique choice of its conjugate form $\H(\t_0 -t)$. 
Note that this conjugation is different from a physical time-reversal operation, under which magnetic field changes sign. The microscopic variables $[\th(t), \phi(t)]$ denote angular
position, and are even functions under time-reversal. 
Thus the time-reversed conjugate trajectory can be denoted as
 $X^\dagger = [\th(\t_0 - t), \phi(\t_0 -t), \H(\t_0 -t)]$. 
The probability of conjugate trajectory   
${\cal P}_- = \prod_{i=1}^N p_i^-$, where 
$p_i^- =    \J_i^- \la \d(\dot \th +\Theta(\t_0 -t)\,)\, \d(\dot \phi + \Phi (\t_0 -t) \ra $, 
where $ \J_i^-$ denotes the relevant Jacobian.  As $\J_i^-=\J_i$, Jacobians drop out of the ratio $p_i^+ / p_i^-$.

Let us outline the calculation of Jacobian in presence of multiplicative noise. 
Given a Langevin dynamics $\dot x = {\cal F}(x) + g(x) \eta(t) $ with multiplicative noise $g(x)\eta(t)$, one may rewrite it as $\dot x/g(x)=h(x) + \eta(t)$ where $h(x)= {\cal F}(x)/g(x)$.
Using a transformation $q(x)$ such that $\dot q = \dot x (dq/dx)$ with $dq/dx = 1/g(x) $,  the Langevin equation may be expressed as $\dot q = h(q) + \eta(t)$, transforming the multiplicative 
noise evolution of $x$ in terms of $q(x)$ evolving under additive noise. 
Then using Stratonovich discretization one can show 
$$\J_i \equiv {\rm det}[\p_{x_i} \eta_i]  = \f{1}{\d t\, g(x_i)} \left[ 1 - \f{\d t}{2} g(x_i) \,\f{\p}{\p{x_i}} \left( \f{{\cal F}(x_i)}{g(x_i)} \right) \right].$$
Note that under time-reversal $x_i$ and as a result ${\cal F}(x_i)$ and $g(x_i)$ remain invariant, and so does $\J_i$.
This behavior is generic for over-damped Langevin equations with multiplicative noise, and remains valid for path probabilities of $(\th,\phi)$ coordinates.

Let us assume that the trajectories considered above describe evolution from initial steady state described by a distribution $P_i(\th_i,\phi_i,\H_i)$ to a final state $P_\ell(\th_\ell,\phi_\ell,\H_\ell)$.
The total probabilities of time-forward  and time-reversed conjugate trajectories are given by ${\cal P}^f [X] = P_i {\cal P}_+$ and ${\cal P}^b [X^\dagger] = P_\ell {\cal P}_-$ respectively, where $X$ and
$X^\dagger$ denote the forward and conjugate processes. 
The Kullback-Leibler divergence of these probabilities 
\bea
D({\cal P}^f || {\cal P}^b) = \sum_X  {\cal P}^f [X] \ln \f{{\cal P}^f [X] }{{\cal P}^b [X^\dagger]}. \nn
\eea
is a non-negative quantity and is a good candidate for the expression of average total EP $\la \D s_t \ra$.
The change in stochastic entropy of the system $\D s = s_\ell - s_i = \kb \ln (P_i/P_\ell)$.
This is a state function and depends on the exact initial and final micro-states. 
Similar to $\D s$, one can define the total entropy change $\D s_t =\kb \ln  \f{{\cal P}^f [X] }{{\cal P}^b [X^\dagger]}  = \D s + \D s_r$, where $\D s_r = \kb \ln \f{{\cal P}_+}{{\cal P}_- } $ 
is the change in entropy in the reservoir~\cite{Kurchan1998}.  
The above relation for $\D s_t$
readily leads to the integral fluctuation theorem (IFT)~\cite{Seifert2012}, $\la e^{-\D s_t/\kb} \ra = 1$. Note that in deriving IFT, $\sum_X \equiv \sum_X^\dagger$ is used, as the 
Jacobian of transformation from time-forward path $X$ to time-reversed path $X^\dagger$ is unity~\cite{Spinney2012}.
As a result of IFT, and the Jensen inequality one obtains  $\la \D s_t \ra \geq 0$, the second law of thermodynamics. This is
equivalent to the statement  $\dot S_t = d \la s_t \ra / dt > 0$ derived above using the Fokker-Planck equation describing the irreversibility of the non-equilibrium dynamics.

After some algebra, it is possible to show that 
the ratio of two probabilities of forward and reverse paths $\f{{\cal P}_+}{{\cal P}_- } = \exp(\D s_r/\kb)$, where 
\bea
\f{\D s_r}{\kb} &=& V \f{m}{\kb T} \int_0^{\t_0} dt  \le[  (H_\th \dot\th + H_\phi \sin\th \dot\phi) \ri. \nn\\
                        && \le. + \f{1}{m \eta \g} (H_\th \sin \th \dot \phi - H_\phi \dot \th)\ri].
\label{dsr}                                        
\eea
In the last step we used $D_0 = \eta \kb T/V$. 
The definition $\D s_r$ in Eq.(\ref{dsr}) directly leads to the expression of EP in the reservoir $\dot s_r$ in Eq.(\ref{srdot}) derived from Fokker-Planck equation.

Further, in a steady state the total entropy change $\D s_t$ along a time-forward path $\D s^f_t(X)$ is equal and opposite to that along the time-reversed path, $\D s^b_t(X^\dagger) = -\D s^f_t(X)$. 
Using this, one obtains the following detailed fluctuation theorem (DFT)~\cite{Kurchan2007,Crooks1999} 
\bea
\r(\D s_t) &=& e^{\D s_t/\kb} \rho(-\D s_t).
\eea

As is already mentioned, it is possible to consider conjugate trajectories, in various other manner, e.g., considering $X^\dagger = [-\m(\t_0-t), -\H(\t_0-t)]$, or $X^\dagger = [\m(\t_0-t), -\H(\t_0-t)]$, or 
$X^\dagger = [-\m(\t_0-t), \H(\t_0-t)]$ (see Appendix~\ref{Ap-A}). As can be shown  easily, each one of these consideration will give IFT and DFT, but with expressions of { entropy} in reservoir
different from that in Eq.(\ref{dsr}). 

\section{Time-independent uniaxial field and detailed balance}
In presence of an uniaxial external field, the potential energy per unit volume $G(\th)= -H\, m\, \cos\th$, i.e., $H_\phi=0$ as $\p_\phi G =0$.
Assuming the same uniaxial symmetry in probability distribution  $P(\th,t)$ independent of $\phi$, 
$\p_\phi P=0$, leads to  $\p_\phi J_\phi=0$ (see Eq.(\ref{current})\,).
Thus  the Fokker-Planck equation reduces to
\bea
\p_t P(\th,t) = (\sin \th)^{-1} \p_\th (\sin \th\, J_\th ).
\eea
The detailed balance condition
requires vanishing of the dissipative current $J_\th=0$ leading to the 
canonical Boltzmann distribution $P = P_0 \exp [-G(\th)/\kb T]$,
which still allows for the presence of a divergence-less  current [Eq.(\ref{current})] in the
azimuthal direction $J_\phi = -g' (\p_\th G) P(\th) $~\cite{Jr1963} (see Fig.\ref{eqm}($b$)).  
A torque due to $\H$ acting on the magnetization $\m$ leads to precessional probability current $J_\phi $ along $\phi$, and to EP. The situation is similar to 
a particle in a harmonic trap under constant external torque, thereby,  producing entropy~\cite{Chernyak2006,Tome2006}.
The state under constant $\H$ is characterized by $J_\th =0$, $J_\phi(\th) \neq 0$ controlled solely by the 
Boltzmann distribution $P(\th)$.

The change in system entropy between initial state $P_i(\th_i)$ to final state $P_{\ell}(\th_\ell)$ is
\bea
\f{\D s}{\kb} =  [G(\th_\ell) - G(\th_i)] = -H m [\cos\th_\ell - \cos\th_i].
\label{ds}
\eea 
Let us write down the expression of $\D s_r$, for uniaxial time-independent magnetic field such that $H_\phi=0$. 
Thus the expression simplifies to
\bea
\f{\D s_r}{\kb} 
&=& V \f{m}{\kb T} \int_0^\t dt \le[ H_\th \dot \th + \f{1}{m \eta \g} H_\th \sin\th \, \dot \phi\ri].
\label{dsr_1}
\eea

\section{Stochastic energy balance}
The rate of stochastic energy gain per unit volume $\dot G  = -\H \cdot \dot\m - \m \cdot \dot\H$.  
In this expression, the rate of work done by the magnetic field is $\dot W= -\m \cdot \dot\H$. 
The stochastic energy balance is given by $\dot G = \dot q + \dot W $ where stochastic heat absorption by the
system $\dot q = -\H \cdot \dot\m$. 
In the spherical polar coordinate, the stochastic heat absorption per unit volume can be expressed as
\bea
\dot q = -\H\cdot \dot\m  &=& - [\hat\th H_\th  + \hat\phi H_\phi ] \cdot [\hat\th\, m \dot\th + \hat\phi\, m \sin\th \dot\phi]  \nn\\
&=&  - m \le[ H_\th \dot\th  + H_\phi \sin\th \,\dot\phi \ri] \, .
\label{heat_1}
\eea

The above result does not give $\dot Q = V \dot q$ which will obey $\dot Q = - T \dot s_r$. 
The expression of the discrepancy between $-\dot Q/T$ and $\dot s_r$ (see Eq.(\ref{srdot})) is the excess EP, and is given by the quantity $(V/T)[( H_\th \sin \th\, \dot \phi - H_\phi \dot \th )/\eta\g]$. 
Note that, even for uniaxial time-independent external field with $H_\phi=0$,   the trajectory average 
$J_\phi$, which depends on $\sin\th\,\dot\phi$,
remains non-zero,  although, the system equilibrates in the sub-space $\th$. 
The excess EP disappears if the motion of the macro-spin is restricted to two dimensions, as then the role of external field on the magnetization becomes equivalent to a tangential external 
force acting on a diffusing particle moving on the circumference of a circle (see Appendix-\ref{Ap-B}).

Using the Fokker-Planck equation we have shown that the specific form of $\dot s_r$ derived there, leads to total EP $\dot S_t \geq 0$, consistent with the second law of thermodynamics.
One may  redefine the rate of heat dissipation as $ - \dot Q \equiv  T \dot s_r$, independent of how it is to be split into the rate of work done, and the rate of change in internal energy~\cite{Seifert2012}.
Thus it is possible to rewrite $\dot q$ and $\dot W$ keeping $\dot G = \dot q + \dot W$ intact, such that
$\dot Q \equiv - T \dot s_r$ is obeyed. At this point, note that Eq.(\ref{srdot}) gives the form 
\bea
T \dot s_r &=& V \le [ \H\cdot \dot \m  + \f{1}{\eta \g} \le( H_\th \sin \th \dot \phi - H_\phi \dot \th \ri) \ri]\, .
\eea
Thus the redefined quantities will  have the following forms:
\bea
\dot q &\equiv& -\H \cdot \dot\m - \f{1}{\eta \g} \le( H_\th \sin \th \dot \phi - H_\phi \dot \th \ri ),  \label{heat_2}\\
\dot W &\equiv& - \m \cdot \dot\H + \f{1}{\eta \g} \le( H_\th \sin \th \dot \phi - H_\phi \dot \th \ri ).
\label{work_2}
\eea
The redefined rate of work done has excess contribution from angular motion due to spin torque. 
In the context of changing one equilibrium state to another, such redefinition is disadvantageous. As the change in the corresponding free energy  can not  be related to quasi-static work done~\cite{Seifert2012}. However, the current system is intrinsically out of equilibrium, and such guiding principle is not strictly applicable.

Note that in the presence of a time-independent uniaxial magnetic field the system remains out of equilibrium, and  the work done on the system due to spin torque should be dissipated as heat with 
$\la \dot Q \ra = \f{H V}{\eta \g} \la \sin^2\th \, \dot \phi \ra$.
Positivity of total entropy production requires this spin torque contribution in entropy. 

Recently we performed a separate study of EP~\cite{Bandopadhyay2014} 
using a generalized Langevin dynamics of spins, where, unlike in the current study, the spin amplitude is allowed to fluctuate~\cite{Ma2012}.  
This Langevin equation does not involve multiplicative noise, unlike Eq.(\ref{LLG1}).  Our calculation showed that  the  excess EP is related to the behavior of 
phase space variables under time reversal, and does not depend on the constraint of constant $m$ imposed by the LLG equation.

\section{Distribution of entropy production}
\label{simu}
We numerically evaluate the distribution of total EP $\D s_t = \D s + \D s_r$ over trajectories of various durations $\t_0$, in presence of 
a time-independent magnetic field $H$ along $z$-direction. 
This is done by integrating Eq.(\ref{llg2}), expressing the magnetization in units of $m$,  energy in units of  $\kb T$, and  using $H = \kb T/m$. 
{In numerical integration, we used a stochastic generalization of Heun scheme~\cite{Garcia-Palacios1998}. This method is known to converge to the solutions of stochastic differential 
equations interpreted in the Stratonovich sense. In simulations, we used time step $\d t = 0.001 \t$, where $\t=m/\g \kb T$ sets the unit of time.}

\begin{figure}[t] 
\begin{center}
\includegraphics[width=8cm]{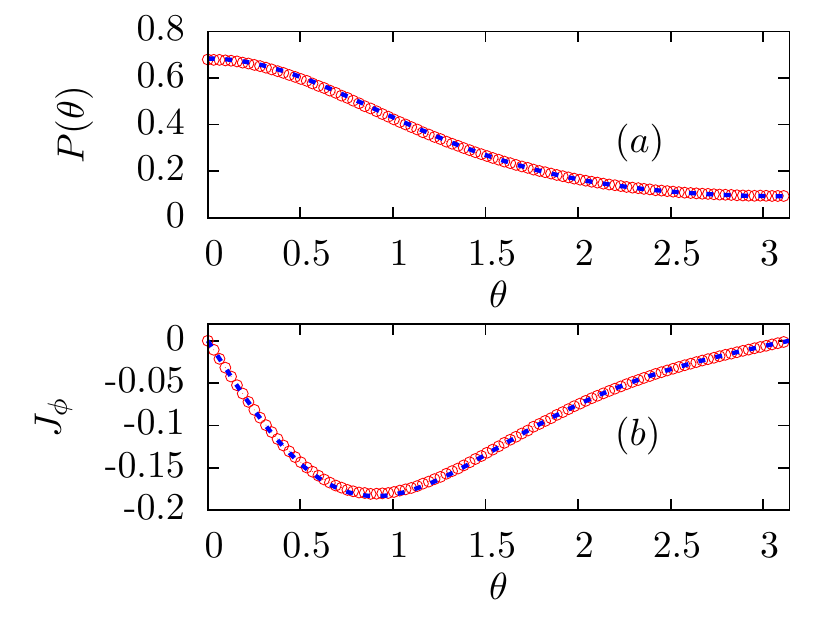}
\caption{(Color online) The data points show simulation results and the dashed lines denote plots of analytic expressions. 
In both the figures $\th$ is expressed in radians. 
($a$) Equilibrium probability distribution $P(\th)$ obtained from numerical integration of Eq.(\ref{llg2}) of magnetization amplitude $m=1$ at 
$H \hat z$ with $H=1$, and $\kb T=1$. This is compared with the expression $P_0 \exp(\cos\th)$ describing the equilibrium distribution. 
($b$)~Azimuthal current  $J_\phi$ as a function of polar angle $\th$. 
}
\label{eqm}
\end{center}
\end{figure}

To test the validity of our numerical integration, we first obtain the { equilibrium} distribution $P(\th)$ that agrees with  
analytical form $P_0 \exp(-G(\th)/\kb T)$ with $G(\th) = -m H \cos\th$ (Fig.\ref{eqm}($a$)). The steady state
supports a probability current $J_\phi(\th)$ in the azimuthal direction. A typical steady state trajectory is shown in Fig.\ref{conf}.
Fig.\ref{eqm}($b$) shows the polar dependence of azimuthal current obtained from simulations, and its comparison 
with the theoretical expression 
\beq
J_\phi = g' \, m H_\th\, P(\th). \nn
\eeq

\begin{figure}[t] 
\begin{center}
\includegraphics[width=8cm]{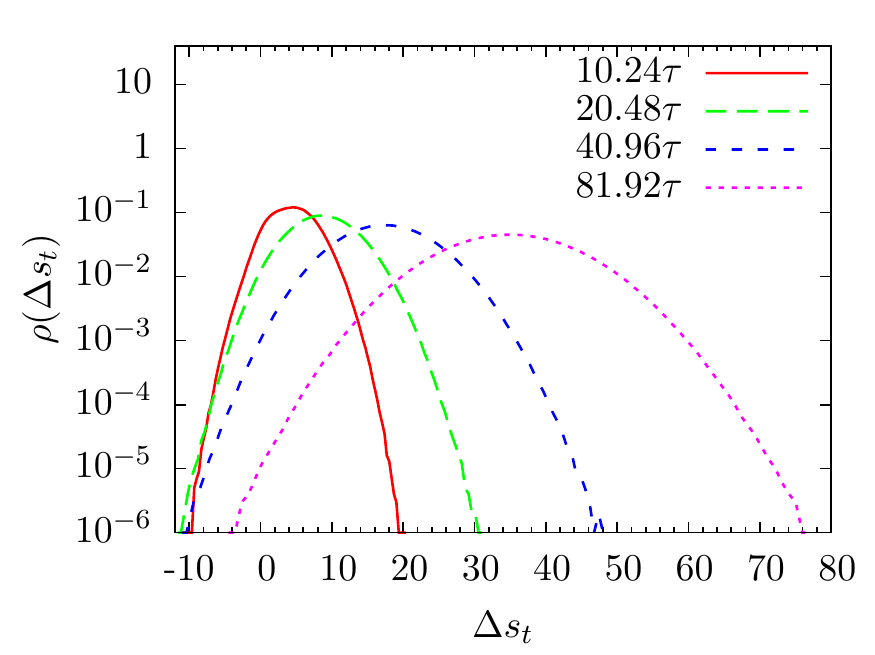}
\caption{(Color online) Probability distribution of total EP $\r (\D s_t)$ calculated in the presence of a constant external field $H \hat z$ with $H=1$.
The calculations are performed after collecting data over $\t_0 = 10.24, 20.48, 40.96, 81.92 \t$. 
}
\label{Pentropy}
\end{center}
\end{figure}

\begin{figure}[t] 
\begin{center}
\includegraphics[width=8cm]{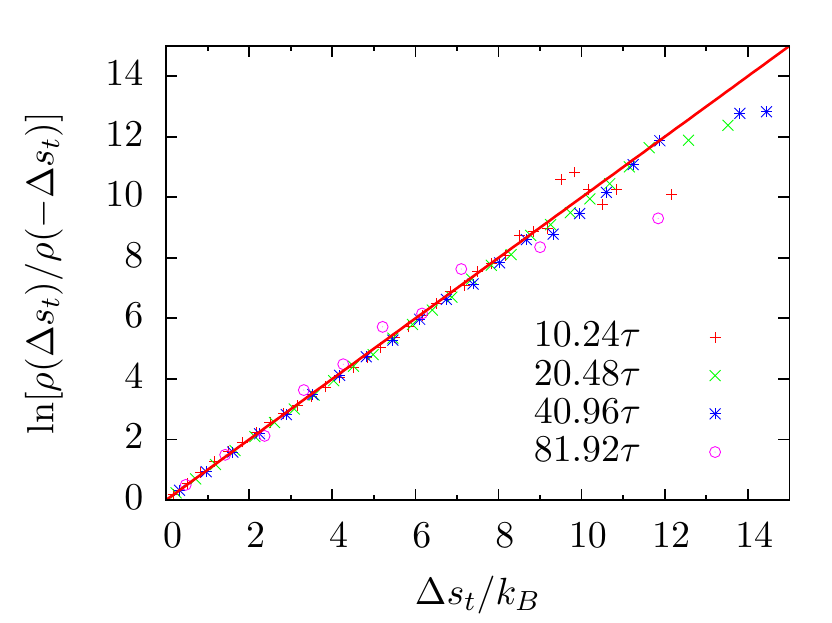}
\caption{(Color online) Ratio of probability distributions of positive and negative EP, $\r(\D s_t)/ \r(-\D s_t)$ calculated from the data 
shown in Fig.~\ref{Pentropy}. The solid line is a plot of the function $\D s_t/\kb$. 
}
\label{PSratio}
\end{center}
\end{figure}

In Fig.~\ref{Pentropy} we show the probability distributions of EP $\r(\D s_t)$ 
calculated numerically using $\D s_t=\D s + \D s_r$, and expressions of $\D s$ and $\D s_r$ from Eq.s  (\ref{ds}) and (\ref{dsr_1}) respectively. 
The distributions are calculated after collecting data over $10^7$ realizations for various durations of ${\t_0}$ as indicated in Fig.~\ref{Pentropy}. 
Appreciable probability of negative EP is clearly visible. 
With increase in ${\t_0}$, the distribution broadens and the peak position shifts towards higher values of entropy. 
From each $\r(\D s_t)$  curve, we obtain the  ratio of probabilities of positive and negative entropy productions, $\r(\D s_t)/\r(- \D s_t)$. 
As is shown in Fig.~\ref{PSratio}, this ratio shows good agreement with the DFT, $\r(\D s_t)/\r(- \D s_t) = \exp(\D s_t/\kb)$. 
The deviation of data from the analytic function 
at $\D s_t/\kb \gtrsim 10$ is due  to poor statistics at large negative entropies.


Note that the probability distribution $\r(\D s_t)$ that obeys DFT, was obtained from time evolution in forward direction, without any need to define a conjugate trajectory under time-reversal.
This again shows that the property of distribution function  $\r(\D s_t)$ is encoded in the dynamics. The choice of conjugate trajectories used for deriving fluctuation theorems have to be
consistent with the expression of entropy production obtained from the Fokker-Planck equation, governing the dynamics of probability distribution of micro-states.  
This requirement restricts the choice of conjugate trajectories to obey specific symmetry, which we presented in the main text. We discuss other
possible choices of conjugate trajectories in the Appendix-\ref{Ap-A}. This is to demonstrate that, although, the corresponding entropy-like quantities obey DFT and IFT, all of them 
fail to capture the physical entropy production.
For time independent $H$ that generates a directed spin-torque, the only contribution to 
EP in the environment comes from the azimuthal dynamics. The fact that the simulated probability distribution $\r(\D s_t)$ obeys DFT, means $\la \D s_t \ra \geq 0$  with $\D s_t$ containing the excess EP due to gyroscopic motion.

\section{Outlook}
We derived fluctuation theorems involving entropy production (EP) in a macrospin undergoing stochastic rotational dynamics in presence of external magnetic field.
The fluctuation theorems were derived using ratio of probabilities of time-forward and reversed trajectories.
While it is possible to choose the reversed trajectories in various manner, we argued that the choice needs to be consistent with the Fokker-Planck equation.
The constraint of constant amplitude of magnetization renders a dynamics that naturally involves a multiplicative noise. 
The magnetic field generates a spin torque, even when the field is time-independent,  driving the system out of equilibrium. 
This led to an excess EP in the environment, which appears to be inconsistent with the expression of heat dissipated derived from energy conservation. 
However this anomaly can be lifted by redefining the expressions of dissipated heat and stochastic work done.
Using numerical simulations, we obtained the distributions of EP over various time intervals, and showed that they agree with the detailed fluctuation theorem.

The steady state of a macrospin dynamics under time-independent magnetic field behaves differently in two subspaces -- while 
the axial current is zero leading to a Boltzmann distribution, azimuthal current remains non-zero. 
The corresponding heat dissipation may be measured by calorimetry. On the other hand,  whole trajectories $[\th(t),\phi(t)]$
may be followed using  Kerr microscopy~\cite{Hiebert2003,Kruglyak2005}.
These two independent measurements may be used to check experimentally whether Eq.(\ref{heat_1}) or Eq.(\ref{heat_2}) describe the actual heat dissipation.
Such measurements would improve our understanding of the relationship between stochastic heat dissipation and EP due to torque. 
Similar situation arises, e.g., for a particle in harmonic trap under constant external torque, producing entropy~\cite{Chernyak2006,Tome2006}.
We hope that our work will elicit further discussions on this technologically relevant and fundamentally important topic. 

\acknowledgments
We thank Udo Seifert, Abhishek Dhar, Punyabrata Pradhan, and Abhishek Chaudhuri  for valuable discussions. AMJ thanks DST, India for financial support.

\appendix
\section{Alternative choices of conjugate trajectories}
\label{Ap-A}
We present three more choices of conjugate trajectories and their consequences. 
Denoting the path probability of reverse trajectory  
$X^\dagger = [-\m(\t_0-t), -\H(\t_0-t)]$ by ${\cal P}_-^{(1)}$, one can show that the 
ratio ${\cal P}_+ / {\cal P}_-^{(1)} = \exp(\D s_r^{(1)}/\kb)$ where 
\bea
\f{\D s_r^{(1)}}{\kb} = V\f{m}{\kb T} \int_0^{\t_0} dt \le[ H_\th \dot \th + H_\phi  \sin\th\, \dot \phi \ri] = -\f{\D Q}{T}.
\label{dsr1}
\eea
This implies that $\D s_t^{(1)} = \D s + \D s_r^{(1)}$ obeys the IFT, and as a result $\la \D s_t^{(1)} \ra \geq 0$. The EP under the time
reversal symmetry considered here is associated with dissipative components of probability currents under the same symmetry~\cite{Bandopadhyay2014}. As the external driving $\H$ in the present case  is 
not symmetric under time reversal, the DFT will have the form $\r(\D s_t^{(1)}) = e^{\D s_t^{(1)}/\kb} \rho^\dagger(-\D s_t^{(1)})$ where $\rho^\dagger$
denotes the probability calculated along the conjugate trajectory.

For the choice of conjugate trajectory in which $\H$ alone changes sign, such that the probability of conjugate trajectory $X^\dagger = [\m(\t_0-t), -\H(\t_0-t)]$ is denoted by ${\cal P}_-^{(2)}$, one obtains the 
ratio ${\cal P}_+ / {\cal P}_-^{(2)} = \exp(\D s_r^{(2)}/\kb)$ where
\bea
\f{\D s_r^{(2)}}{\kb} = V\f{m}{\kb T} \int_0^{\t_0} dt \f{1}{m \eta \g}\le[ H_\th \sin\th\, \dot \phi - H_\phi {\dot\th} \ri].
\label{dsr2}
\eea
Again, $\D s_t^{(2)} = \D s + \D s_r^{(2)}$ obeys the IFT and DFT. It is interesting to note that EP in the environment as shown in the main text $\D s_r = \D s_r^{(1)} + \D s_r^{(2)}$ [see Eq.(\ref{dsr})]. 

The third alternative is to consider conjugate trajectories in which $\m$ alone changes sign, i.e., $X^\dagger = [-\m(\t_0-t), \H(\t_0-t)]$. Denoting the probability of conjugate trajectory ${\cal P}_-^{(3)}$,
one obtains ${\cal P}_+ / {\cal P}_-^{(3)} = \exp(\D s_r^{(3)}/\kb)$, with  $\D s_r^{(3)}=0$. By construction, $\D s_t^{(3)} = \D s$ also obeys the IFT and DFT.

\section{Stochastic thermodynamics: Spin confined to 2D}
\label{Ap-B}
Here we consider that the stochastic spin-rotation is confined to two dimensions (2D). The Langevin dynamics is described by 
\bea
\f{d \m}{dt} = \g\le[ H_0 \hat z + \h(t) - \eta \f{d\m}{dt}\ri] \times \m
\eea
where we assumed an external field perpendicular to this 2D plane. If the spin is restricted to rotate on the $(m,\phi)$ plane, $d\m/dt = m \dot \phi \hat \phi$ with $\dot \phi = d \phi/dt$. 
In a strictly two-dimensional dynamics, neglecting the out of plane motion due to Gilbert damping, 
the equation of motion simplifies to 
\bea
\dot \phi = \g [H_0 + h_z(t)], 
\eea
with $\la h_z(t) \ra=0$, $\la h_z(t) h_z(t') \ra = 2 D_0 \d(t-t')$.
This is equivalent to the over damped Langevin motion in 1D $\dot x = \mu [ \xi(t) + f ]$ with mobility $\mu$, Gaussian white noise $\xi(t)$, and external force $f$.
Thus one obtains a stochastic version of  first law of thermodynamics $\D q + \D W = 0$ with heat absorbed by the system 
$\D q = \int dt \dot \phi [-\dot \phi/\g + h_z(t) ]$ and work done $\D W = \int dt \dot \phi H_0$.

Let us assume that the initial and final micro-states are described by $\phi_i$ and $\phi_\ell$ respectively. We consider time evolution starting from a single micro-state picked up from the initial distribution $P_i(\phi_i)$, which evolves to one of the final micro-states obeying a distribution $P_\ell(\phi_\ell)$. The probability of a time-forward  path evolved from $t=0$ to $\t$ is described by the stochastic field
$h_z(t) = \dot\phi/\g - H_0$ with ${\cal P}_+ \propto \exp [-(1/4D_0) \int_0^\t dt\, h_z^2(t)]$ where $D_0=\kb T/\g$. Under time reversal the stochastic noise is described by $[-\dot\phi/\g - H_0]$. Thus the ratio of these path-probabilities are  given by ${\cal P}_+/{\cal P}_- = \exp [ (1/D_0 \g ) \int_0^\t dt H_0 \dot \phi ] $. Now using the definition of work done and the first law derived above, we may rewrite the relation as
${\cal P}_+/{\cal P}_- = \exp [ -(1/D_0 \g )\D q]$. This quantity accounts for the entropy change in the reservoir ${\cal P}_+/{\cal P}_- = \exp(\D s_r/\kb)$~\cite{Seifert2005}, with
$\D s_r = - \kb \D q/(D_0 \g)= - \D q/T $. Thus the 2D counter part of the full 3D dynamics does not have any discrepancy in terms of the definition of dissipated heat with EP in the 
reservoir, unlike in 3D as described in the main text. The anomaly in EP and subsequent resolution of it via redefinition of heat and work is required in 3D, as the system 
maintains non-equilibrium rotational current in azimuthal direction even when the external magnetic field is time-independent.

\bibliographystyle{prsty}

\end{document}